\documentclass[aps, preprint,superscriptaddress,showkeys,floatfix,amsmath,amssymb,prx,onecolumn,11pt,a4paper
]{revtex4-2}
\usepackage[T1]{fontenc}
\usepackage[utf8]{inputenc}
\usepackage{graphicx}
\usepackage{hyperref}
\hypersetup{
    colorlinks=true,
    linkcolor=blue,
    citecolor=red,
    filecolor=magenta,
    urlcolor=cyan}

\renewcommand{\selectlanguage}[1]{} 

\begin{document}

\title{Probing individual phonon-polaritonic nanoparticle-on-mirror cavities by infrared nanospectroscopy}

\author{Isabel Pascual Robledo}
\thanks{These authors contributed equally to this work.}
\affiliation{CIC nanoGUNE BRTA, Tolosa Hiribidea 76, 20018 Donostia-San Sebasti\'an, Spain}
\affiliation{Materials Physics Center, CSIC-UPV/EHU, Manuel de Lardizabal 5, 20018 Donostia-San Sebasti\'an, Spain}
\affiliation{Department of Physics, University of the Basque Country UPV/EHU, Barrio Sarriena s/n, 48940 Leioa, Spain}

\author{Iker Herrero Le\'on}
\thanks{These authors contributed equally to this work.}
\affiliation{CIC nanoGUNE BRTA, Tolosa Hiribidea 76, 20018 Donostia-San Sebasti\'an, Spain}
\affiliation{Department of Physics, University of the Basque Country UPV/EHU, Barrio Sarriena s/n, 48940 Leioa, Spain}

\author{Karol Ko{\l}\k{a}taj}
\affiliation{Department of Physics, University of Fribourg, Chem. du Mus\'ee 3, 1700 Fribourg, Switzerland}

\author{Guillermo P. Acuna}
\affiliation{Department of Physics, University of Fribourg, Chem. du Mus\'ee 3, 1700 Fribourg, Switzerland}

\author{Javier Aizpurua}
\affiliation{Donostia International Physics Center (DIPC), Paseo Manuel de Lardizabal 4, 20018 Donostia-San Sebasti\'an, Spain}
\affiliation{Department of Electricity and Electronics, FCT-ZTF, University of the Basque Country EHU, Barrio Sarriena z/g, 48940 Leioa, Spain}
\affiliation{IKERBASQUE, Basque Foundation for Science, 48009 Bilbao, Spain}

\author{Philippe Roelli}
\email{p.roelli@nanogune.eu}
\affiliation{CIC nanoGUNE BRTA, Tolosa Hiribidea 76, 20018 Donostia-San Sebasti\'an, Spain}

\author{Rainer Hillenbrand}
\email{r.hillenbrand@nanogune.eu}
\affiliation{CIC nanoGUNE BRTA, Tolosa Hiribidea 76, 20018 Donostia-San Sebasti\'an, Spain}
\affiliation{Department of Electricity and Electronics, FCT-ZTF, University of the Basque Country EHU, Barrio Sarriena z/g, 48940 Leioa, Spain}
\affiliation{IKERBASQUE, Basque Foundation for Science, 48009 Bilbao, Spain}

\date{\today}

\begin{abstract}
Nanoparticle-on-mirror (NPoM) cavities enable extreme light confinement
and strong light--matter interactions, but their realization with
phonon-polariton materials in the mid-infrared spectral range remains
largely unexplored. Here, we use nano-FTIR spectroscopy to study the
near-field response of individual phononic NPoM cavities formed by gold
nanoparticles on a quartz substrate supporting phonon-polaritons. By
placing a metal tip on top of the NPoM and recording the tip-scattered
field, we observe two reproducible cavity resonances, identified as the
fundamental and a second-order antenna modes by numerical simulations.
The calculations show that, in absence of the tip, the NPoM cavity
exhibits ultrasmall mode volumes (\(V\) \(\sim\)
10\textsuperscript{3} nm$^{3}$) and high quality factors
(\(Q\) \(\sim\) 100), resulting in extraordinary field
intensity enhancements (\(F\) \(\sim\) 10\textsuperscript{4})
and Purcell factors (\(P_{F}\) \(\sim\) 10\textsuperscript{9}).
They also indicate that the nano-FTIR tip enables efficient excitation
and readout of these phononic NPoM modes without perturbing them, while
enhancing the intrinsic local field intensity in the NPoM gap by two
orders of magnitude. Our results establish phononic NPoM cavities as a
promising platform for mid-infrared nanophotonics and pave the way for
ultrasensitive vibrational spectroscopy based on nano-FTIR measurements
of individual cavities.
\end{abstract}


\maketitle

\section{Introduction}

Nanoparticle-on-mirror (NPoM) cavities typically consist of a plasmonic
metallic nanoparticle, commonly gold or silver, positioned above a
metallic mirror and separated by a nanometric gap.\cite{mmock20081,mbaumberg2}
Together, the nanoparticle (NP) and mirror form a well-defined optical
nanocavity in which the nanoparticle plasmon couples with its mirror
image, forming plasmon polaritons in the gap. As a result, light is
confined to volumes well below the diffraction limit, leading to a
strong field confinement and enhancement in the nanoparticle--mirror
gap.\cite{mcirac20123,mchen20184} These cavities are readily fabricated through
chemical synthesis and are generally more practical to realize than
other lithographically fabricated plasmonic cavities.\cite{mzhu20145}

NPoM cavities support highly tunable visible (VIS) resonances with
ultrasmall normalized mode volumes (\(V_{\mathrm{norm}} \sim
10^{-6} \ \lambda^{3} -
10^{-5} \ \lambda^{3}\), where \(\lambda\) is the incident
wavelength) and moderate quality factors (\(Q \sim 10 -
50\)), yielding large Purcell enhancements (\(P_{F} \sim
10^{5} - 10^{6}\)).\cite{mbaumberg2,mroelli6}
Consequently, NPoMs have become a versatile platform for fluorescence
enhancement,\cite{mhoang20167} surface-enhanced Raman scattering
(SERS),\cite{mbenz20168,mlu20249,mpeng10} nanoscale sensing,\cite{mcarnegie202011,mchen202112}
nonlinear optics\cite{mli202113,mroelli202514} and plasmon-driven
chemistry.\cite{mjeong202415,moksenberg202116} Intriguingly, picocavities may emerge
within the NPoM gap,\cite{mpark201017,murbieta18,mcarnegie19} further concentrating the
electromagnetic field to atomic-scale dimensions, thereby enabling
single-molecule spectroscopy and strong coupling.\cite{mbaumberg202220}

By contrast, mid-infrared (mid-IR) resonant NPoM cavities remain
comparatively unexplored despite their potential for vibrational
molecular spectroscopy in this spectral range. Coupling molecular
vibrations to the strongly confined NPoM fields could enable
surface-enhanced infrared absorption spectroscopy (SEIRA) and
vibrational strong coupling (VSC) at the nanoscale. While conventional
metallic nanoparticles do not typically support plasmonic resonances in
the mid-IR, doped semiconductor particles can be used in this context to
support resonant modes in this spectral range.\cite{mnaik21}
However, the performance of such materials is often limited by their
higher optical losses and reduced oscillator strength, leading to
broader resonances and weaker field confinement compared to metallic
plasmonic systems\cite{mkhurgin201822} and
metasurfaces.\cite{mniemann202423,mweber202324}

Phonon-polaritonic nanostructures based on polar dielectrics provide an
alternative platform for mid-IR nanophotonics. Polar materials such as
silicon carbide (SiC), hexagonal boron nitride (hBN), or molybdenum
trioxide ($\alpha$-MoO\textsubscript{3}) support low-loss surface and volume
phonon polaritons within their Reststrahlen bands, enabling strong
infrared field confinement with significantly reduced losses compared to
conventional plasmonic systems.\cite{mhillenbrand200225,mcaldwell26,mgaliffi27,mlow28,mduan202329,mtamagnone30} Phonon-polariton
nanoresonators have therefore emerged as promising candidates for
applications including SEIRA\cite{mautore201831} and
VSC.\cite{mdolado202232}

Although strong infrared field enhancement and confinement may be
realized in NPoM cavities using polar crystals, preparing nanoparticles
from polar crystals with controlled shapes and sizes remains
challenging.\cite{mwang33} A practical alternative can take
advantage of the combination of non-resonant metallic particles with a
polar dielectric substrate. Recently silver (Ag) nanocubes deposited on
a SiC substrate were shown to support confined phonon-polaritonic NPoM
modes through the excitation of SiC surface phonon polaritons by the NP
near field.\cite{mklein202534} These cavities reported normalized mode
volumes as small as \(V_{\mathrm{norm}}\) \(\sim\)
10\textsuperscript{-8} \(\lambda^{3}\) and quality factors of \(Q\)
\(\sim\) 170. However, the extremely small extinction
cross-section of these systems restricts conventional spectroscopy to
ensemble measurements, preventing direct access to the response of
individual cavities.

Probing single phonon-polaritonic NPoM cavities is nevertheless
essential for understanding cavity-to-cavity variations and the
influence of local geometry and material properties on the optical
cavity response. Probing individual NPoM cavities could also enable
studies of intrinsic emitter decay channels and nanoscale mid-IR
light-matter interactions, including VSC with very small molecular
ensembles. In the visible spectral range, single particle
spectroscopy\cite{mmock20081,mlei201235,mhu201036} has become a key tool for
observing dynamical reshaping of the cavities and the formation of
atomic protrusions, which enable enhanced interaction between light and
single emitters.\cite{mcarnegie202011,mchen202112,mchikkaraddy201637}

A powerful method for probing individual nanostructures in the infrared
is nanoscale Fourier transform infrared (nano-FTIR)
spectroscopy.\cite{mhuth201138,mhillenbrand39} In nano-FTIR, an atomic force
microscope (AFM) tip serves as a nanoscale optical antenna that
concentrates broadband infrared radiation at its apex, generating a
strong near-field enhancement. When the sample is brought into close
proximity to the tip apex, the localized near field interacts with the
sample, which leads to a modification of the tip-scattered light
depending on the local optical properties of the sample. By
interferometrically detecting the scattered light as a function of
frequency, nano-FTIR provides infrared amplitude and phase spectra with
nanoscale spatial resolution determined by the AFM tip radius.

Here, we experimentally demonstrate that infrared spectra of individual
phonon-polaritonic NPoMs can be measured using nano-FTIR spectroscopy.
Simulations indicate that positioning the tip above the NP enhances the
electromagnetic field amplitude inside the nanoparticle--mirror gap by a
factor of \(\sim\) 20. This enhancement occurs without
significantly perturbing the properties of the cavity modes, including
their spectral positions, quality factors, and spatial extent of the
electric near-field distribution. This approach enables the experimental
study of mid-IR nanocavity physics at the level of individual NPoMs,
with potential applications in nanoscale infrared sensing and VSC.

\section{Results and discussion}

\begin{figure}[htbp]
\centering
\includegraphics[width=0.95\textwidth]{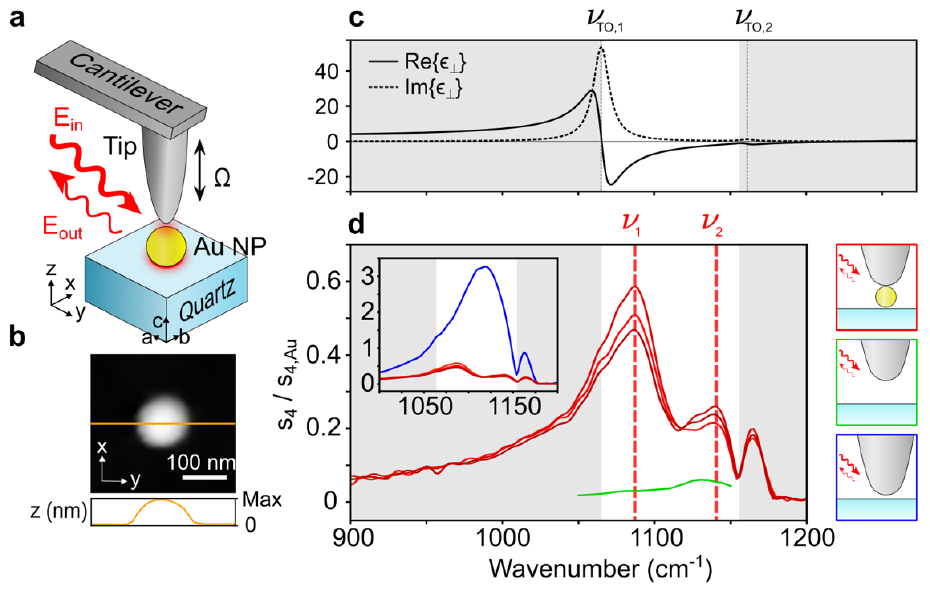}
\caption{\textbf{Nano-FTIR spectroscopy on a
phonon-polaritonic NPoM cavity.} (a) Schematics of the experiment. A
PtIr tip with 50 nm apex radius oscillates at a frequency \(\Omega\)
above a AuNP deposited on a c-cut oriented quartz mirror. The red arrows
indicate the incident illumination \(E_{\mathrm{in}}\) and the backscattered
light \(E_{\mathrm{out}}\). (b) Topography image (top) and line profile (bottom)
of a 60 nm AuNP on quartz. (c) Real (solid line) and imaginary (dashed
line) parts of the in-plane dielectric component
\(\varepsilon_{\bot}(\nu)\) of quartz, fitted from far-field
reflectivity measurements.\(\ \nu_{TO,1} =\) 1065 cm$^{-1}$ and
\(\nu_{TO,2} =\) 1160.7 cm$^{-1}$ mark the TO phonon frequencies of quartz.
The white area marks the spectral region where
\(\operatorname{Re}\{\varepsilon_{\bot}\}\) is negative and monotonic. (d) nano-FTIR
amplitude spectra, \(s_{4}(\nu)/s_{4,\mathrm{Au}}(\nu)\), of three NPoM cavities
(red curves), of the quartz mirror when the oscillating tip is retracted
by 60 nm (green curve), and of the quartz mirror when the oscillating
tip is in contact (blue curve in the inset). Schematics on the right
illustrate the different sample configurations. Vertical dashed red
lines mark the peak spectral positions of the NPoM nano-FTIR amplitude
spectra.}
\label{fig:main1}
\end{figure}

\hyperref[fig:main1]{Figure~1a} illustrates nano-FTIR spectroscopy of an individual phononic
NPoM cavity. The NPoM sample is prepared by drop-casting a colloidal
solution of Au nanoparticles (AuNPs) onto a quartz substrate acting as a
mirror (see \hyperref[si:section-8-experimental-methods]{Experimental Methods} in the Supplement Information, SI),
resulting in the formation of NPoM cavities. A representative topography
image of an individual NPoM is shown in \hyperref[fig:main1]{Fig.~1b}. Each AuNP is coated
with a \(\sim\) 1 nm thin dielectric layer of citrate formed
during NP synthesis, which defines a nanoscale gap between the particle
and the mirror. Quartz is chosen as a mirror because its transverse
optical (TO) phonons give rise to a spectral window where surface phonon
polaritons (SPhPs) can be excited. In addition, quartz can be polished
to achieve atomically flat surfaces with sub-nanometer roughness, which
is essential for achieving NPoM cavities with ultra-small mode volumes
comparable to those reported for plasmonic NPoMs.

As a uniaxial anisotropic crystal, quartz is characterized by its
in-plane \(\varepsilon_{\bot}\) and out-of-plane
\(\varepsilon_{\parallel}\) dielectric tensor
components.\cite{mgervais40,mwinta201941,mtappert201342} We use c-cut samples and determine
the in-plane component \(\varepsilon_{\bot}\) (\hyperref[fig:main1]{Fig.~1c}) by fitting
far-field reflectivity measurements (see \hyperref[si:section-1-far-field-reflectivity-measurements-on-quartz]{SI Section 1}). Two phonons are
observed at \(\nu_{TO,1} =\) 1065 cm$^{-1}$ and \(\nu_{TO,2} =\) 1160.7
cm$^{-1}$, giving rise to a spectral region where
\(\operatorname{Re}\{\varepsilon_{\bot}\}\) \textless{} 0, which enables the excitation
of SPhPs. We note that \(\operatorname{Re}\{\varepsilon_{\bot}\}\) is non-monotonic
near \(\nu_{TO,2}\), which complicates the analysis of NPoM cavity
modes. We therefore focus our study on the spectral range between
\(\nu_{TO,1}\) and \(\nu_{TO,2}\), where \(\operatorname{Re}\{\varepsilon_{\bot}\}\)
\textless{} 0 and behaves monotonically (white region in \hyperref[fig:main1]{Fig.~1c}). 

To obtain infrared near-field spectra of individual NPoM cavities, we
position the metal tip of the nano-FTIR setup on top of the
corresponding AuNP and illuminate it with broadband mid-infrared laser
radiation (\hyperref[si:section-8-experimental-methods]{Experimental Methods}, SI). Due to the lightning rod effect,
the tip concentrates the incident mid-infrared radiation into a
nanoscale spot at the tip apex, which illuminates the individual NPoM
cavity and excites its modes. The near-field response of the NPoM cavity
modifies the field back-scattered by the tip. By recording the
tip-scattered field with an asymmetric Fourier-transform spectrometer,
we obtain near-field amplitude and phase spectra. To isolate the
near-field response from the background scattering, the tip oscillates
normal to the quartz mirror at a frequency \(\Omega\), and the detector
signal is demodulated at a higher frequency \(n\Omega\) (\(n =\) 1, 2,
3, \ldots). The demodulated signal yields background-free near-field
spectra, of which we show and analyze only the amplitude spectra,
denoted \(s_{n}(\nu)\).

\begin{figure}[htbp]
\centering
\includegraphics[width=0.95\textwidth]{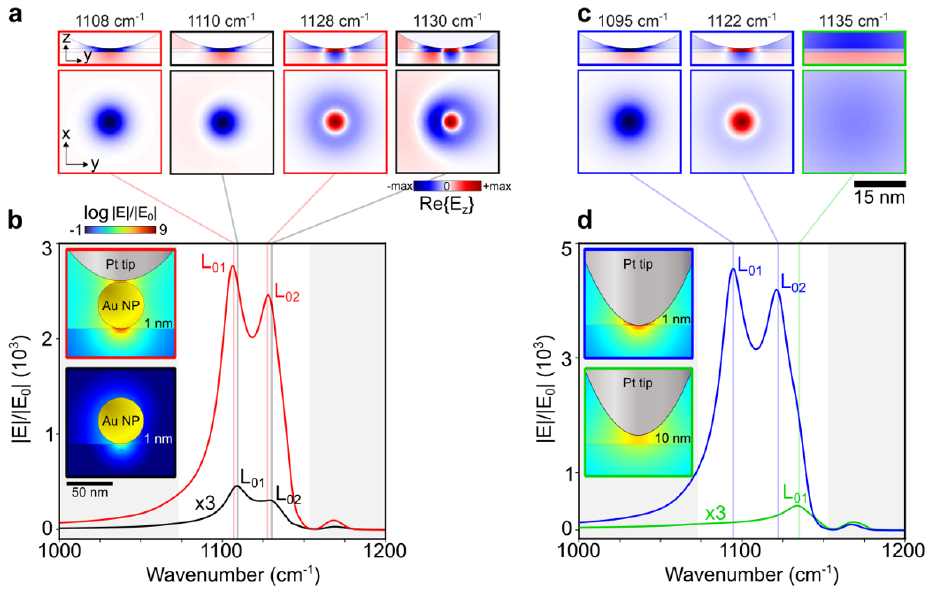}
\caption{\textbf{Numerical simulations of spectral field
enhancement and mode profiles in phonon-polaritonic nanocavities.} (a)
Vertical (top) and mid-gap horizontal (bottom) near-field distribution
(real part of the out-of-plane electric field, \(\operatorname{Re}\{ E_{z}\}\)) in the
NPoM cavity gap (50 nm AuNP on quartz) at the wavenumbers indicated
above. Red frames: static Pt tip (50\,nm radius) positioned 2 nm above
the NPoM cavity; black frames: NPoM cavity without tip. (b) Spectra of
the maximum field enhancement, \(\max\{|E|(\nu)/|E_{0}|(\nu)\}\), in the
mid-gap horizontal plane of a NPoM cavity (50 nm AuNP on quartz). Red
spectrum: static Pt tip (50\,nm radius) positioned 2 nm above NPoM
cavity; black curve: NPoM cavity without tip. Insets show vertical
near-field distribution at 1110 cm$^{-1}$. (c) Vertical
(top) and horizontal (bottom) near-field distributions
(\(\operatorname{Re}\{ E_{z}\}\), in the midgap plane) at indicated wavenumbers for a
static Pt tip (25\,nm radius) positioned 1\,nm (blue frames) and 10\,nm
(green frame) above a quartz surface (no AuNP present). (d) Spectra of
the maximum field enhancement \(\max\{|E|(\nu)/|E_{0}|(\nu)\}\) in the
mid-gap horizontal plane for a static Pt tip (25\,nm radius) at 1\,nm
(blue frames) and 10\,nm (green frames) above the quartz. Insets show
vertical near-field distribution at the frequency of the \(\mathrm{L}_{01}\)
mode. In all calculations, the illumination is a p-polarized wave of
amplitude \(|E_{0}| = \) 1 V/m, and incident $35^\circ$ relative to the quartz
surface plane.}
\label{fig:main2}
\end{figure}

\hyperref[fig:main1]{Figure~1d} shows nano-FTIR amplitude spectra recorded on three individual
60 nm diameter NPs (red), measured with a PtIr tip of 50 nm apex radius
and normalized to a 200 nm thick gold reference,
\(s_{4}(\nu)/s_{4,\mathrm{Au}}(\nu)\). Each spectrum shows two peaks at
approximately \(\nu_{1} =\) 1090 cm$^{-1}$ and \(\nu_{2} =\) 1140 cm$^{-1}$
(dashed lines in \hyperref[fig:main1]{Fig.~1d}) with the same relative peak heights for all
three particles, demonstrating a reproducible spectral response. We
attribute these peaks to the NPoM cavity resonances, as corroborated by
numerical simulations in \hyperref[fig:main2]{Fig.~2}.

To gain further experimental insight into the origin of these peaks, we
record nano-FTIR spectra in the absence of the AuNP. In a first
measurement, the oscillating tip is in direct contact with the quartz
surface, yielding a broad peak (blue curve in the inset of \hyperref[fig:main1]{Fig.~1d}) with
a higher near-field amplitude signal compared to that measured when the
tip is positioned above the AuNP (red curves in \hyperref[fig:main1]{Fig.~1d}). This spectral
response arises from tip-excited SPhPs in quartz, which lead to a
resonant near-field coupling between the tip and quartz surface,
analogue to the coupling that forms the phonon-polaritonic modes within
the NPoM gap.\cite{mautore201943,mamarie201144} When the tip is retracted by a
distance comparable to the diameter of the AuNP (details in \hyperref[si:section-2-spectra-as-a-function-of-tipquartz-separation]{SI Section
2}), we again observe a single broad peak, but with reduced amplitude,
which arises from the weaker near-field coupling in the increased
tip--quartz gap (green curve in \hyperref[fig:main1]{Fig.~1d}).\cite{mamarie201144} These
measurements confirm that the two peaks observed when the tip is
positioned above the AuNP (red curves in \hyperref[fig:main1]{Fig.~1d}) arise from the
presence of the AuNP and not from direct tip--quartz near-field
coupling. Therefore, these peaks can be attributed to phonon-polaritonic
resonances at the NPoM cavity.

To corroborate that phonon-polaritonic NPoM cavity modes are responsible
for the two peaks observed in the nano-FTIR spectra (red curves in \hyperref[fig:main1]{Fig.
1d}), we perform electromagnetic numerical simulations of the phononic
NPoM, with a 1 nm gap, in the presence of the nano-FTIR tip (details in
see \hyperref[si:section-9-numerical-methods]{SI Numerical Methods} and \hyperref[si:section-3-effect-of-quartz-anisotropy-on-the-local-field-enhancement-in-the-npom-cavity]{SI Section 3}). We compute the near-field
distribution around the NPoM cavity, \(|E(x,y,z;\ \nu)|/|E_{0}(\nu)|\),
over a range of wavenumbers $\nu$ between 1000 cm$^{-1}$ and 1200 cm$^{-1}$, where
\(|E|\) is the amplitude of the local electric field and \(|E_{0}|\) is
the incident field amplitude. We consider two configurations: (i) a NPoM
cavity formed by a 50 nm diameter AuNP, and (ii) the same NPoM cavity
with a 50 nm apex radius tip positioned 2 nm above the AuNP. As an
example, the insets of \hyperref[fig:main2]{Fig.~2b} show the near-field distribution
\(|E(y,z;\ \nu)|/|E_{0}(\nu)|\) at \(\nu\) \(\sim\) 1110 cm$^{-1}$ for the two
configurations, with the tip (top inset) and without the tip (bottom
inset). In both cases, we observe strong near-field confinement and
enhancement in the gap between the AuNP and the quartz mirror. When the
tip is on top of the NPoM, this field enhancement is significantly
stronger, indicating that the near field at the tip apex provides
enhanced local illumination of the nanoparticle and thereby stronger
excitation of NPoM cavity modes.

We first analyze the spectral response of the first configuration, that
is the NPoM cavity without tip. To that end, we plot in \hyperref[fig:main2]{Fig.~2b} the
maximum field enhancement, \(\max\{|E|(\nu)/|E_{0}|(\nu)\}\), in the
horizontal mid-gap plane as a function of wavenumber \(\nu\) (black
curve). We observe two peaks in the spectral region of interest (white
area in \hyperref[fig:main2]{Fig.~2b}), at 1110 cm$^{-1}$ and 1130
cm$^{-1}$, indicating phonon-polaritonic NPoM cavity
resonances. To identify the NPoM modes corresponding to these
resonances, we analyze the near-field distributions around the
nanoparticle--mirror gap (without tip) at each resonance wavenumber
(black-framed maps in \hyperref[fig:main2]{Fig.~2a}). The top panel displays vertical
cross-sections through the AuNP and quartz mirror (\(y - z\) plane),
while the bottom panel shows horizontal cross-sections across the
mid-gap (\(x - y\) plane). In both cases, we plot the real part of the
\(z\)-component of the electric field, \(\operatorname{Re}\{ E_{z}\}\). At 1110
cm$^{-1}$, the field distribution exhibits a single maximum
at the gap center in both vertical and horizontal cross-sections,
revealing that this resonance peak corresponds to the fundamental
antenna mode of the NPoM cavity, denoted \(\mathrm{L}_{01}\).\cite{mzhang201945}
Here, \(\mathrm{L}\) denotes the longitudinal (dipolar) character of the mode
along the vertical axis between the nanoparticle and the mirror. The
first index specifies the azimuthal mode order while the second index
represents the radial mode order. These numbers correspond to the number
of electric field amplitude \(|E|\ \)maxima along the angular coordinate
and the radial direction respectively. At 1130 cm$^{-1}$,
the horizontal cross-section exhibits two concentric maxima around the
gap center, characteristic of a second-order axially symmetric
mode.\cite{mtserkezis201546} We identify this mode as the second-order
antenna mode \(\mathrm{L}_{02}\), based on the charge distribution on the NP
surface (\hyperref[si:section-4-surface-charge-distribution-of-the-npom]{SI Section 4}). We note that both mode patterns deviate from
perfect circular symmetry, due to the tilted illumination.

After identifying the cavity modes, we calculate the mode volumes and
quality factors for both of them (see \hyperref[si:section-5-mode-volume-and-quality-factor-calculation]{SI Section 5}). For the fundamental
\(\mathrm{L}_{01}\) mode, we obtain a mode volume of \(V =\) 1309
nm$^{3}$ (\(V_{\mathrm{norm}} =\) 1.8 \(\cdot\) 10\textsuperscript{-9}
\(\ \lambda^{3}\)) and a quality factor \(Q =\) 81, corresponding to a
Purcell Factor of \(P_{F} =\) 3.4 \(\cdot\) 10\textsuperscript{9}. For the
second-order \(\mathrm{L}_{02}\) mode, we obtain \(V = \) 690
nm$^{3}$ (\(V_{\mathrm{norm}} = \) 1.0 \(\cdot\) 10\textsuperscript{-9}
\(\ \lambda^{3}\)), \(Q = \) 96 and \(P_{F} = \) 7.3 \(\cdot\)
10\textsuperscript{9}. Notably, these phononic NPoM cavities exhibit
higher quality factors and similar mode volumes compared to plasmonic
NPoMs in the visible range.\cite{mroelli6}

As observed in \hyperref[fig:main2]{Fig.~2a,b}, the presence of the tip does not significantly
modify the NPoM spectral response, with resonance positions, field
distribution, and quality factors remaining essentially unchanged (\(Q
=\) 80 and \(Q = \) 108, respectively). The tip, however, strongly enhances
the local electric field in the gap to a factor of
\(|E(\nu)|/|E_{0}(\nu)|\) \(\sim\) 2700, corresponding to a local intensity
enhancement of \(F\) \(\sim\) 7 \(\cdot\) 10\textsuperscript{7}. This enhancement
occurs without modifying the lateral extent of the mode, which is about
8 nm, by more than 3\% (see \hyperref[si:section-6-modification-of-the-modes-upon-tip-approach]{SI Section 6}).

Comparing these simulation results with the experimental near-field
amplitude spectra obtained from the NPoM cavities (red curves, \hyperref[fig:main1]{Fig.~1d}),
we find good qualitative agreement. In particular, the presence of two
resonance peaks in the spectral region of interest is properly
reproduced, although in the experiment these peaks are broader and
appear at lower frequencies. These discrepancies may originate from
geometrical parameters not included in the simulation, such as the
precise optical thickness of the NP capping layer, the likely presence
of a NP facet, the optical anisotropy of the quartz substrate, or
defects and strain from polishing of the quartz surface. Despite these
quantitative differences, the simulations reproduce the experimentally
observed spectral features and allow us to identify the peaks of the red
spectra in \hyperref[fig:main1]{Fig.~1d} as the \(\mathrm{L}_{01}\) and \(\mathrm{L}_{02}\) NPoM cavity modes.

For reference and comparison with the experimental near-field spectra
recorded directly on quartz (blue and green curves in \hyperref[fig:main1]{Fig.~1d}), we also
calculate the near-field distributions and the field enhancement in the
gap between a static tip of 25 nm apex radius (same as the AuNP particle
radius) and a quartz mirror, following the same procedure as for the
NPoM simulations shown in \hyperref[fig:main2]{Fig.~2a,b} (\hyperref[si:section-9-numerical-methods]{Numerical Methods}, SI). When the
static tip is positioned 1\,nm above the quartz surface, the calculated
spectrum (blue curve in \hyperref[fig:main2]{Fig.~2d}) exhibits two peaks, identified as the
\(\mathrm{L}_{01}\) and \(\mathrm{L}_{02}\) modes, based on the near-field distributions
(blue-framed maps in \hyperref[fig:main2]{Fig.~2c}). This shows that the tip and quartz form a
phonon-polaritonic cavity, qualitatively similar to an NPoM cavity.
However, the modes in the tip--quartz gap exhibit larger field
enhancement and are red-shifted compared to those in the NPoM
configuration. This can be explained by the larger polarizability of the
Pt tip compared to the AuNP, which leads to a stronger near-field
interaction in the gap created between the tip and the quartz substrate.
Increasing the tip--quartz distance to 10 nm, strongly reduces the
field enhancement in the gap (green curve in \hyperref[fig:main2]{Fig.~2d}). In addition, only
one peak is observed in this situation, corresponding to the \(\mathrm{L}_{01}\)
mode and significantly blueshifted. This behavior contrasts with that
observed for the NPoM configuration (\hyperref[fig:main2]{Fig.~2a,b}), where the gap defining
the NPoM cavity remains fixed independently of the presence of the tip
and its position.

By comparing the calculated spectra in \hyperref[fig:main2]{Fig.~2d} with the experimental
near-field amplitude spectrum (blue curve in \hyperref[fig:main1]{Fig.~1d}, inset), a clear
discrepancy is observed. In the experiment, only a single broad peak is
present, whereas the simulation with a static tip positioned 1 nm above
the quartz surface (blue curve in \hyperref[fig:main2]{Fig.~2d}) exhibits two distinct peaks
corresponding to the cavity modes. On the other hand, the experimental
spectrum is reproduced by calculations of the tip-scattered field using
an analytical model (\hyperref[si:section-7-calculated-demodulated-near-field-spectra-on-quartz]{SI Section 7}), which includes tip oscillation and
signal demodulation.\cite{mvoronin202547} We explain the appearance of a
single broad peak in both experimental and calculated near-field
amplitude spectra, rather than two distinct cavity resonances, by the
tip oscillation. This leads to a modulation of the tip--mirror gap that
alters the near-field interaction throughout the oscillation cycle,
thereby modifying both the resonance frequencies and field enhancements
of the cavity modes (\hyperref[si:section-7-calculated-demodulated-near-field-spectra-on-quartz]{SI Section 7}). The averaging inherent to the signal
demodulation process consequently merges the two resonances into a
single broadened spectral feature.

Summarizing the simulations and their comparison with experiments, both
the NPoM cavity and the tip--quartz system support phonon-polaritonic
cavity modes. However, in nano-FTIR measurements on bare quartz, the
oscillating tip introduces a time-dependent tip--sample coupling, which,
together with signal demodulation, leads to a broad and shifted peak in
the nano-FTIR spectra. In contrast, in nano-FTIR measurements of the
NPoM cavity, the particle--quartz gap is maintained fixed during the tip
oscillation cycle, such that the NPoM cavity modes remain well defined.
Consequently, the peaks observed in the nano-FTIR spectrum remain at the
same spectral positions. In the tip-on-NPoM configuration, the tip
primarily provides enhanced local field excitation without altering the
NPoM cavity. This makes this configuration particularly well suited for
implementation of stable and controlled field-enhanced infrared
spectroscopy of ultrathin layers. 

\begin{figure}[htbp]
\centering
\includegraphics[width=0.95\textwidth]{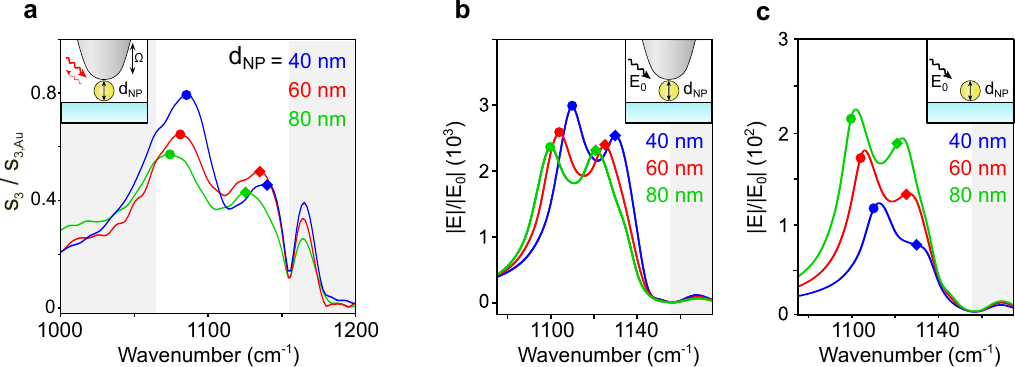}
\caption{\textbf{Nano-FTIR spectroscopy on phononic NPoMs of
different size.} (a) nano-FTIR amplitude spectra, \(s_{3}/s_{3,\mathrm{Au}}\), of
NPoM cavities with NPs of diameter \(d_{\mathrm{NP}} = \) 40 nm (blue curve), 60
nm (red curve) and 80 nm (green curve). Schematics on the top-left
illustrate the variation of \(d_{\mathrm{NP}}\). (b) Spectra of the maximum field
enhancement, \(\max\{|E|(\nu)/|E_{0}|(\nu)\}\), in the mid-gap
horizontal plane of NPoM cavities (AuNPs on quartz) of diameter
\(d_{\mathrm{NP}} = \) 40 nm (blue curve), 60 nm (red curve) and 80 nm (green
curve). In all spectra a Pt tip (50\,nm radius) is positioned 2 nm above
the NPoM cavity. (c) Spectra of \(\max\{|E|(\nu)/|E_{0}|(\nu)\}\), in
the mid-gap horizontal plane of NPoM cavities (AuNPs on quartz) of
diameter \(d_{\mathrm{NP}} = \) 40 nm (blue curve), 60 nm (red curve) and 80 nm
(green curve) without the tip. In all panels symbols mark the positions
of the \(\mathrm{L}_{01}\) (dots) and \(\mathrm{L}_{02}\) (diamonds) NPoM cavity modes.}
\label{fig:main3}
\end{figure}

We explore now the possibilities to tune the phononic NPoM modes with NP
size. To that end we show in \hyperref[fig:main3]{Fig.~3a} nano-FTIR spectra of phononic NPoMs
with AuNPs of diameters \(d_{\mathrm{NP}} = \) 40 nm (blue curve), \(d_{\mathrm{NP}} = \) 60
nm (red curve) and \(d_{\mathrm{NP}} = \) 80 nm (green curve). All spectra exhibit
two peaks, in agreement with the experimental and theoretical analysis
discussed in \hyperref[fig:main1]{Fig.~1} and \hyperref[fig:main2]{Fig.~2}. Based on our previous analysis, we
identify the peaks to be the \(\mathrm{L}_{01}\) and \(\mathrm{L}_{02}\) NPoM modes,
marked by dots and diamonds in \hyperref[fig:main3]{Fig.~3a}, respectively. Notably, both
peaks redshift with increasing particle size, accompanied by an overall
reduction in signal amplitude. This trend is reproduced by numerical
calculations of \(\max\{|E|(\nu)/|E_{0}|(\nu)\}\) at the center of the
NPoM gap in the presence of the tip (\hyperref[fig:main3]{Fig.~3b}), confirming that phononic
NPoM cavity modes can be tuned via nanoparticle size within the
Reststrahlen band of the substrate. We note that similar size-dependent
spectral shifts have been reported in plasmonic NPoM
systems.\cite{mbenz20168}

In addition to the spectral redshift, the simulations including the tip
(\hyperref[fig:main3]{Fig.~3b}) show a reduction of the local field enhancement for larger
NPs. This contrasts with plasmonic NPoMs, where the local field
enhancement typically increases with the NP size.\cite{mbenz20168} To
elucidate whether this reduction arises from the phononic NPoM itself or
from the presence of the tip, we performed simulations of phononic NPoMs
without the tip (\hyperref[fig:main3]{Fig.~3c}). In this case,
\(\max\{|E|(\nu)/|E_{0}|(\nu)\}\) increases with NP size, analogue to
the tendency in plasmonic NPoMs. This demonstrates that the reduction of
the local field enhancement with increasing nanoparticle size (\hyperref[fig:main3]{Fig.~3b})
does not originate from the NPoM cavity itself but rather from the
presence of the tip. This effect can be attributed to the increased
tip--quartz separation with increasing NP size, which weakens the
electromagnetic coupling between the tip and quartz substrate, as
discussed in \hyperref[fig:main1]{Fig.~1} and \hyperref[fig:main2]{Fig.~2}.

The strong influence of the tip on how the field enhancement scales with
nanoparticle size raises the question of whether it also affects the
peak positions and the observed size-dependent redshift. Comparing
simulations with and without the tip (\hyperref[fig:main3]{Figs.~3b and 3c}), we find that the
peak positions exhibit the same size-dependent shift in both cases,
confirming that the observed redshift with increasing particle size is
intrinsic to the NPoM cavity rather than tip-induced. This conclusion is
further supported by previous studies reporting that increasing the
tip--substrate distance induces a blueshift,\cite{mamarie201144} in
contrast to the redshift observed here.

\section{Conclusion}

In summary, we demonstrate nano-FTIR spectroscopy of individual
phonon-polariton NPoM cavities formed by gold nanoparticles on quartz
substrates. These nanocavities support well-defined mid-infrared
phonon-polaritonic resonances, including the fundamental \(\mathrm{L}_{01}\) and
a second-order \(\mathrm{L}_{02}\) antenna-like NPoM modes, identified via
electromagnetic simulations. The simulated cavities exhibit ultrasmall
mode volumes \(V \sim 700-1300\) nm$^{3}$ (\(V_{\mathrm{norm}}\) \(\sim\)
10\textsuperscript{-9}\(\ \lambda^{3}),\ \)and high quality factors
(\(Q \sim 80-110\)). In addition, the electric field in the phononic
NPoM cavity is strongly enhanced within the nanoscale cavity gap
(\(|E|/|E_{0}|\) \(\sim\) 150) and is further enhanced by the presence of the
tip (\(|E|/|E_{0}|\) \(\sim\) 2700). We thus anticipate the possibility to
engineer strongly enhanced IR light--matter interaction in future
experiments, where the gap is filled with molecules. We further show
that the NPoM cavity resonances can be tuned within the Reststrahlen
band of the phononic substrate by changing the nanoparticle size to
match for example, specific molecular vibrations. Phonon-polariton NPoMs
thus represent a promising platform for nanoscale infrared sensing,
vibrational strong coupling, and mid-infrared nanophotonics.

Importantly, we demonstrate that the nano-FTIR tip acts as a
non-invasive local excitation and detection probe, enabling
single-particle spectroscopy of mid-IR NPoMs. The tip does not perturb
the stable cavity modes but enhances the electromagnetic field in the
NPoM gap by more than one order of magnitude, which may benefit future
nonlinear vibrational studies of minute amounts of matter in the NPoM gap, 
such as vibrational sum-frequency generation\cite{mroelli202514} and
two-dimensional infrared spectroscopy.\cite{msaurabh201649,mxiang202150}

Detailed experimental and numerical methods are provided in the Supplementary Information.

\section*{Acknowledgements}

R.H. and I.H. acknowledge financial support from Grant CEX2020-001038-M
funded by the Spanish MICIU/\allowbreak AEI/\allowbreak 10.13039/\allowbreak 501100011033 and by ERDF/EU;
Grant PID2021-123949OB-I00 funded by the Spanish
MICIU/\allowbreak AEI/\allowbreak 10.13039/\allowbreak 501100011033 and by ERDF/EU. J.A. and I.P.
acknowledge financial support from Grant PID2022-139579NB-I00 funded by
MICIU/\allowbreak AEI/\allowbreak 10.13039/\allowbreak 501100011033 and by ERDF/EU and Grant IT 1526-22 from
the Basque Government for consolidated groups of the Basque University,
through the Department of Science, Universities and Innovation of the
Basque Government. P.~R. acknowledges financial support from the Swiss
National Science Foundation (Grant No.~206926) and from the European
Union's Horizon 2020 research and innovation program under the Marie
Sk{\l}odowska-Curie grant agreement No.~101065661.

R.H. is a co-founder of Neaspec GmbH, which now is a part of Attocube
systems GmbH, a company producing s-SNOM systems, such as the one used in
this study. The remaining authors declare no competing financial
interest.


\clearpage
\begin{center}
\textbf{\large -- Supplementary Information -- \\ Probing individual phonon-polaritonic nanoparticle-on-mirror cavities by infrared nanospectroscopy}
\end{center}
\setcounter{equation}{0}
\setcounter{figure}{0}
\setcounter{table}{0}
\setcounter{section}{0}
\makeatletter
\renewcommand{\thesubsection}{\textbf{Supplementary Section \arabic{subsection}}}
\renewcommand{\figurename}{\textbf{Supplementary Figure}}
\renewcommand{\tablename}{\textbf{Supplementary Table}}
\renewcommand{\theequation}{S\arabic{equation}}
\renewcommand{\thefigure}{S\arabic{figure}}
\renewcommand{\thetable}{S\arabic{table}}
\renewcommand{\theHequation}{SI.\arabic{equation}}
\renewcommand{\theHfigure}{SI.\arabic{figure}}
\renewcommand{\theHtable}{SI.\arabic{table}}
\renewcommand{\theHsection}{SI.\arabic{section}}
\renewcommand{\theHsubsection}{SI.\arabic{subsection}}
\renewcommand*{\citenumfont}[1]{S#1}
\renewcommand*{\bibnumfmt}[1]{[S#1]}

\makeatother

\subsection{Far-field reflectivity measurements on quartz}
\label{si:section-1-far-field-reflectivity-measurements-on-quartz}

\begin{figure}[htbp]
\centering
\includegraphics[width=0.6\textwidth]{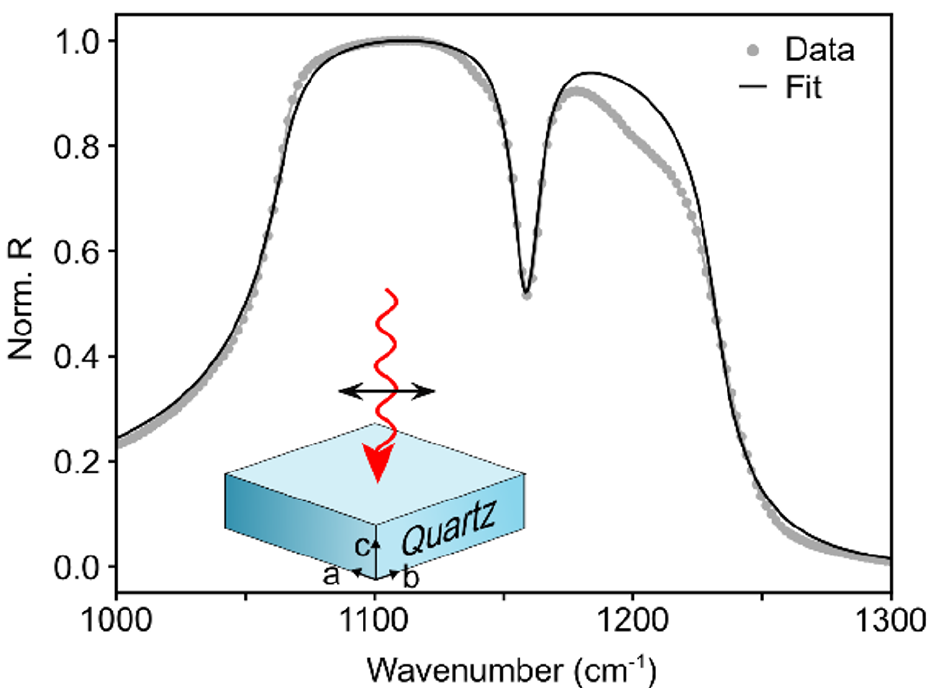}
\caption{Normalized reflectivity spectrum of
quartz under normal incidence for a c-cut orientation (see schematics).
The dots represent the experimental data, and the solid line corresponds
to the fit obtained using a double Lorentz oscillator model.}
\label{fig:siS1}
\end{figure}

To determine the in-plane dielectric function \(\varepsilon_{\bot}\) of
quartz presented in the main text (\hyperref[fig:main1]{Fig.~1c}) and used in all the
numerical calculations, we perform far-field FTIR reflectivity
measurements (dots in \hyperref[fig:siS1]{Fig.~S1}). The measurements are carried out at near
normal incidence on a c-cut quartz crystal (see schematics in \hyperref[fig:siS1]{Fig.~S1}).
In this configuration, the measured optical response corresponds to the
in-plane dielectric function \(\varepsilon_{\bot}\),\cite{swinta20191}
which is obtained through the normalized reflectivity \(R(\nu)\)
\begin{equation}
R(\nu) = \frac{\mid n_{\bot}(\nu) - 1 \mid^{2}}{\mid n_{\bot}(\nu) + 1 \mid^{2}}
\label{eq:S1}
\end{equation}
where \(n_{\bot}(\nu) = \sqrt{\varepsilon_{\bot}(\nu)}\) is the
complex-valued refractive index corresponding to the in-plane dielectric
function. In the spectral range of interest, the in-plane dielectric
response of quartz is dominated by two infrared-active phonon
modes.\cite{swinta20191,sgervais2} Accordingly, we model the in-plane
dielectric function \(\varepsilon_{\bot}\) using a double Lorentz
oscillator model to account for these two phonon modes:
\begin{equation}
\varepsilon_{\bot}\ (\nu) = \varepsilon_{\bot\infty}\ \left( 1 + \frac{\nu_{LO,1}^{2} - \ \nu_{TO,1}^{2}}{\nu_{TO,1}^{2} - \nu^{2} - \ i\ \gamma_{1}\ \nu} + \frac{\nu_{LO,2}^{2} - \ \nu_{TO,2}^{2}}{\nu_{TO,2}^{2} - \nu^{2} - \ i\ \gamma_{2}\ \nu} \right)
\label{eq:S2}
\end{equation}
where \(\varepsilon_{\bot\infty}\) is the high-frequency permittivity.
\(\nu_{TO,j}^{\ }\), \(\nu_{LO,j}^{\ }\) and \(\gamma_{j}\) (\(j = \) 1,
2) are the transverse and longitudinal optical phonon frequencies, and
the damping constant of each phonon mode, respectively.

Fitting the experimental reflectivity with Eq.~\ref{eq:S1} using the dielectric
model of Eq.~\ref{eq:S2} yields the solid curve shown in \hyperref[fig:siS1]{Fig.~S1}. The extracted
parameters are\(\ \varepsilon_{\bot\infty}\)= 1.95;
\(\nu_{LO,1}^{\ }\ \)= 1225 cm\textsuperscript{-1}, \(\nu_{TO,1}^{\ }\)
= 1065 cm\textsuperscript{-1}, \(\gamma_{1} = \) 12.5
cm\textsuperscript{-1}; \(\nu_{LO,2}^{\ } = \) 1163.5
cm\textsuperscript{-1}, \(\nu_{TO,2}^{\ } = \) 1160.7
cm\textsuperscript{-1}, \(\gamma_{2}\ \)= 10.5 cm\textsuperscript{-1}.
These parameters are used in the main text to describe the optical
response of the quartz samples employed in the experiments and in all
the numerical calculations performed in this study.

\newpage

\subsection{Spectra as a function of tip--quartz separation}
\label{si:section-2-spectra-as-a-function-of-tipquartz-separation}

In the discussion of \hyperref[fig:main1]{Fig.~1} of the main text, we present spectra
measured on various phononic NPoMs, which show two distinct peaks within
the region of interest. To verify that the observed spectral features
arise from the presence of the NPoM rather than from variations in the
tip--quartz separation, we compare two types of spectra: (i) nano-FTIR
spectra with the tip positioned on top of the NPoM (\hyperref[fig:main1]{Fig.~1d}, red). (ii)
spectrum recorded on quartz at a tip--quartz distance similar to the NP
size, i.e. 60 nm (\hyperref[fig:main1]{Fig.~1d}, green). In this section we explain in detail
how these measurements were performed and the reasons for not using
standard nano-FTIR spectroscopy.

\begin{figure}[htbp]
\centering
\includegraphics[width=0.9\textwidth]{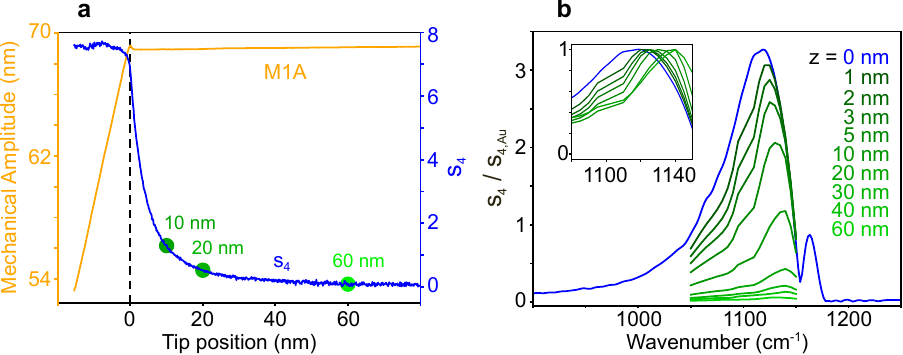}
\caption{\textbf{(a)} Representative retraction curve on
a quartz sample measured at a laser wavenumber of 1095
cm\textsuperscript{-1}, using an oscillating Pt/Ir tip, with tapping
amplitude 70 nm and tip apex radius 50 nm. The mechanical amplitude
recorded by the AFM quadrant detector is shown in orange, while the
fourth order demodulated optical amplitude is shown in blue. The black
dashed line marks the point at which the tip loses contact with the
sample. Green markers indicate the optical amplitude at selected
tip--quartz distances. (b) In blue, nano-FTIR spectrum measured on
quartz. In green, spectra constructed from a set of retraction curves
acquired at different illumination wavenumbers. The optical amplitude is
extracted at fixed tip distances for each wavenumber, as illustrated in
(a). The inset shows the same data after normalization, for improved
visualization of the spectral shift with tip--quartz separation.}
\label{fig:siS2}
\end{figure}

To retract the sample from the tip, the closed-loop feedback controlling
the vertical sample position is switched off, leaving the tip--sample
distance unstable. However, standard nano-FTIR measurements typically
require acquisition times of 30 seconds to a few minutes, making stable
tip--sample positioning essential. Without feedback, the tip position in
our setup is insufficiently stable to record conventional nano-FTIR
spectra in the non-contact regime. To overcome this limitation, we use a
monochromatic quantum cascade laser (QCL) to record retraction curves at
selected wavenumbers, enabling reconstruction of near-field spectra as a
function of the tip--sample distance.

Retraction curves are acquired as follows. The tip is positioned above
the sample and operated in tapping mode under standard feedback
conditions. The feedback controlling the tip--sample distance is then
switched off, and the sample is retracted from the oscillating tip along
the \(z\)-axis using a piezoelectric actuator. During retraction, both
the tip´s oscillation amplitude and the near-field signal are measured
as a function of the tip--sample separation.

An example retraction curve acquired on quartz with the QCL tuned to
1095 cm$^{-1}$ is shown in \hyperref[fig:siS2]{Fig.~S2a}. The mechanical
oscillation amplitude (shown in orange) is used to identify the point at
which the oscillating tip loses contact with the surface. This point,
marked by the dashed line, defines the reference position \(z\ \)= 0.
For positions \(z\ \)\textgreater{} 0, the tip oscillates above the
surface without contact to the sample. When the sample is retracted, i.e
when \(z\) increases, the near-field signal \(s_{4}\) decreases because
of the decreasing near-field interaction between tip and
sample.\cite{samarie20113}

To reconstruct near-field spectra at different tip--quartz separations,
the procedure is repeated at multiple laser wavenumbers spanning 1050 --
1150 cm$^{-1}$. For each wavenumber, a retraction curve is
recorded and the corresponding \(s_{4}(z)\) values are extracted. These
values are then referenced to a standard nano-FTIR spectrum measured on
quartz under identical conditions (same tip, tapping amplitude and
signal demodulation order). By combining the referenced near-field
amplitudes acquired through the QCL tuning range, we obtain spectra as a
function of tip--quartz distance \(z\). The reconstructed spectra are
displayed in \hyperref[fig:siS2]{Fig.~S2b}, showing that the near-field resonance of quartz
redshifts and broadens with increasing tip--sample distance. This
behavior is consistent with previous experimental\cite{samarie20113,shillenbrand20024}
and theoretical studies\cite{srenger20055} on phononic materials.

Comparison of the reconstructed spectra with the nano-FTIR spectra
measured on phononic NPoMs (\hyperref[fig:main1]{Fig.~1d}, red curves) shows that the two
distinct peaks observed in the NPoM spectra within the spectral region
of interest cannot be explained by an increased tip--quartz distance.
Instead, they arise from the presence of the nanoparticle and its
associated phonon-polaritonic cavity modes. 

\newpage

\subsection{Effect of quartz anisotropy on the local field enhancement in the NPoM cavity}
\label{si:section-3-effect-of-quartz-anisotropy-on-the-local-field-enhancement-in-the-npom-cavity}

All the simulations shown in the main text were performed using only the
in-plane dielectric function \(\varepsilon_{\bot}\) of quartz, rather
than its full dielectric tensor
\(\overset{\overleftrightarrow{}}{\varepsilon} = \mathrm{Diag}(\varepsilon_{\bot},\varepsilon_{\bot},\varepsilon_{\parallel})\).\cite{swinta20191} In this section we show that this isotropic
approximation based on \(\varepsilon_{\bot}\) provides an accurate
description of the system, justifying its use in the main text and in
all numerical calculations.

As discussed in the main text, quartz is a uniaxial anisotropic crystal
whose dielectric tensor, for a c-cut orientation, is given by
\(\overset{\overleftrightarrow{}}{\varepsilon} = \mathrm{Diag}(\varepsilon_{\bot},\varepsilon_{\bot},\varepsilon_{\parallel})\).\cite{swinta20191}
We extract the in-plane dielectric function \(\varepsilon_{\bot}\)
directly from FTIR measurements, as discussed in \hyperref[si:section-1-far-field-reflectivity-measurements-on-quartz]{Section S1} of this
Supplementary Information (SI). However, the out-of-plane dielectric
function \(\varepsilon_{\parallel}\ \ \)cannot be directly measured
using our experimental configuration.

\begin{figure}[htbp]
\centering
\includegraphics[width=0.95\textwidth]{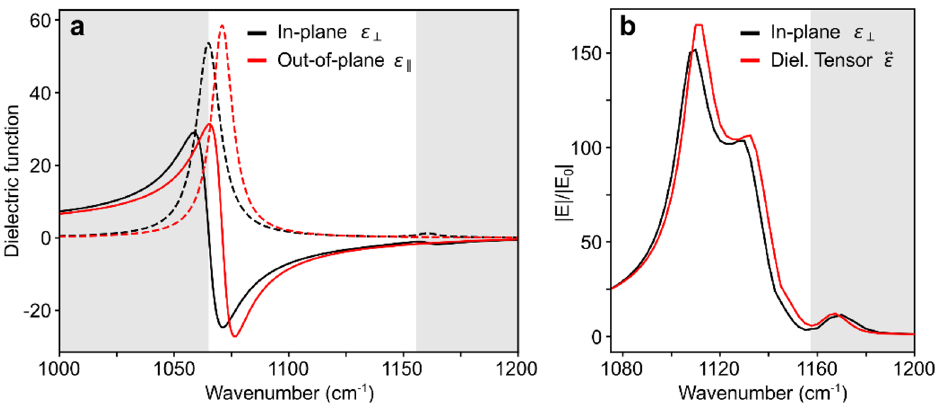}
\caption{\textbf{(a)} In-plane dielectric function
\(\varepsilon_{\bot}\ \)fitted from reflectivity measurements (black)
and estimated out-of-plane dielectric function
\(\varepsilon_{\parallel}\) (red). \textbf{(b)} Maximum electric field
enhancement in the NPoM gap for two cavities with the quartz modelled
as: (i) an isotropic material with dielectric function
\(\varepsilon_{\bot}\) (black), and (ii) an anisotropic material with
dielectric tensor
\(\overset{\overleftrightarrow{}}{\varepsilon} = \mathrm{Diag}(\varepsilon_{\bot},\varepsilon_{\bot},\varepsilon_{\parallel})\)
(red).}
\label{fig:siS3}
\end{figure}

To estimate \(\varepsilon_{\parallel}\), we rely on published FTIR
measurements of quartz from the literature.\cite{swinta20191,sgervais2} These
studies show that \(\varepsilon_{\parallel}\) closely follows the
spectral behavior of \(\varepsilon_{\bot}\). In particular,
\(\varepsilon_{\parallel}\) exhibits a single phonon resonance at 1071
cm$^{-1}$, with a quality factor similar to that of the
1065 cm$^{-1}$ phonon resonance observed for
\(\varepsilon_{\bot}\). The two phonon resonances define a narrow
spectral region (of \(\sim\) 6 cm$^{-1}$) in which
\(\operatorname{Re}\{\varepsilon_{\bot}\}\) \textless{} 0 and
\(\operatorname{Re}\{\varepsilon_{\parallel}\}\) \textgreater{} 0, corresponding to a
region of hyperbolic light propagation. However, the effect of
hyperbolic propagation is expected to be limited due to the following
factors: (i) spectral window is only 6 cm$^{-1}$ wide, (ii)
this window is located approximately 30 cm$^{-1}$ away from
the main NPoM cavity resonances (1100 cm$^{-1}$, as
observed experimentally in \hyperref[fig:main1]{Fig.~1d} and theoretically in \hyperref[fig:main2]{Fig.~2} of the
main text), and (iii) the large negative values of
\(\operatorname{Re}\{\varepsilon_{\bot}\}\) \(\sim -25\) strongly restrict electromagnetic
field penetration into the substrate.

To evaluate the impact of quartz anisotropy, we implement the complete
quartz dielectric tensor
\(\overset{\overleftrightarrow{}}{\varepsilon} = \mathrm{Diag}(\varepsilon_{\bot},\varepsilon_{\bot},\varepsilon_{\parallel})\)
in the numerical simulations. We estimate the out-of-plane dielectric
function \(\varepsilon_{\parallel}\) by modifying the measured in-plane
dielectric function \(\varepsilon_{\bot}\) (black curve in \hyperref[fig:siS3]{Fig.~S3a}). In
particular, we use a dielectric model similar as the one used for
\(\varepsilon_{\parallel}\) in Eq.~\ref{eq:S2}.
\begin{equation}
\varepsilon_{\parallel}(\nu) = \varepsilon_{\parallel \infty}\ \left( 1 + \frac{\nu_{LO,1}^{2} - \ \nu_{TO,1}^{2}}{\nu_{TO,1}^{2} - \nu^{2} - \ i\gamma_{1}\nu} \right),
\label{eq:S3}
\end{equation}

with parameters \(\varepsilon_{\parallel \infty}\)= 1.95,
\(\nu_{LO,1}^{\ }\ \)= 1225 cm\textsuperscript{-1}, \(\nu_{TO,1}^{\ }\)
= 1071 cm\textsuperscript{-1}, \(\gamma_{1}\)= 11.0
cm\textsuperscript{-1}. We plot the resulting dielectric function for
\(\varepsilon_{\parallel}\) as a red curve in \hyperref[fig:siS3]{Fig.~S3a}.

In \hyperref[fig:siS3]{Fig.~S3b} we compare electromagnetic simulations performed using only
\(\varepsilon_{\bot}\), with simulations including the full anisotropic
tensor
\(\overset{\overleftrightarrow{}}{\varepsilon} = \mathrm{Diag}(\varepsilon_{\bot},\varepsilon_{\bot},\varepsilon_{\parallel})\).
In particular, we plot the spectrum of maximum field enhancement in the
NPoM gap \(\max\{|E|(\nu)/|E_{0}|(\nu)\}\). For both types of
simulations, the electric field spectra in the NPoM cavity exhibit two
characteristic peaks, corresponding to the\(\ L_{01}\) and \(\mathrm{L}_{02}\)
modes discussed in the main text. The inclusion of anisotropy only
results in a slight blueshift of the peaks and a modest increase in the
field intensity, consistent with the contribution of the out-of-plane
component \(\varepsilon_{\parallel}\) to the optical response of quartz.

These results indicate that quartz anisotropy has a negligible influence
on the optical response of the NPoM cavities considered in this study.

\newpage

\subsection{Surface charge distribution of the NPoM}
\label{si:section-4-surface-charge-distribution-of-the-npom}

\begin{figure}[htbp]
\centering
\includegraphics[width=0.95\textwidth]{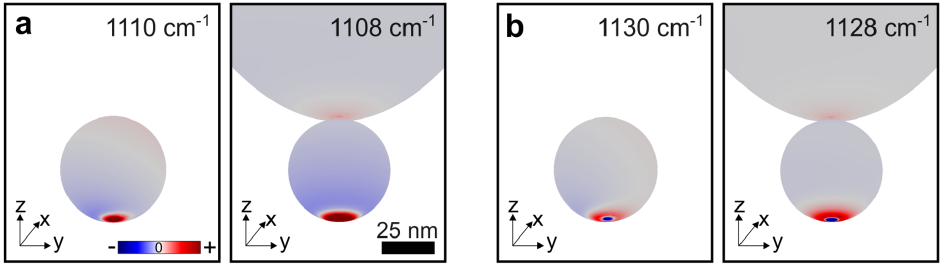}
\caption{\textbf{(a)} Surface charge distribution on the
NP associated with the \(\mathrm{L}_{01}\) NPoM cavity mode in absence (left) and
presence (right) of the tip. \textbf{(b)} Same for the \(\mathrm{L}_{02}\) NPoM cavity
mode. The quartz substrate has been excluded to improve visualization at
the bottom surface of the NP.}
\label{fig:siS4}
\end{figure}

In \hyperref[fig:main2]{Fig.~2} of the main text, the field enhancement spectra in the NPoM
cavity gap exhibit two resonance peaks, both in presence and absence of
the tip (\hyperref[fig:main2]{Fig.~2b}). To identify the modes associated with these
resonances, we plot the nanoparticle surface charge distributions at the
corresponding wavenumbers. For the first resonance, occurring at 1110
cm$^{-1}$ for the bare NPoM (1108 cm$^{-1}$
with the tip), the surface charge distribution exhibits a single sign
change along the nanoparticle´s vertical axis (\hyperref[fig:siS4]{Fig.~S4a}), characteristic
of the fundamental antenna mode, denoted as \(\mathrm{L}_{01}\). The second
resonance, occurring at 1130 cm$^{-1}$ (1128
cm$^{-1}$ respectively), displays two sign changes along
the vertical axis, indicating a second order antenna mode, which we
label as \(\mathrm{L}_{02}\).

\newpage

\subsection{Mode volume and quality factor calculation}
\label{si:section-5-mode-volume-and-quality-factor-calculation}

To quantitatively characterize the cavity modes identified in the main
text (\hyperref[fig:main2]{Fig.~2}), we extract the quality factor \(Q\), the mode volume
\(V\) and the Purcell factor \(P_{F}\). These quantities provide a
quantitative description of the spectral and spatial confinement of the
NPoM cavity modes.

The quality factor \(Q\) is a dimensionless parameter that quantifies
how long the cavity stores energy, corresponding to the ratio between
the energy stored and the energy dissipated per oscillation cycle of the
light inside the cavity. In this context, the definition of \(Q\) is
expressed by the ratio between the resonance wavenumber \(\nu_{j}\ \)and
its linewidth (FWHM) \(\gamma_{j}\), given by
\begin{equation}
Q = \frac{\nu_{j}}{\gamma_{j}}
\label{eq:S4}
\end{equation}
We determine \(Q\) from the spectrum of the field enhancement at the
center of the NPoM cavity under plane-wave excitation
\(\max\{|E|(\nu)/|E_{0}|(\nu)\}\), shown in the \hyperref[fig:main2]{Fig.~2} of the main text.
To extract \(Q\), the amplitude of the electric field in the gap is
fitted using a coherent sum of two uncoupled harmonic oscillators, which
models the response of the electric field enhancement in the gap as:
\begin{equation}
f_{1}(\nu) = \left| {\widetilde{f}}_{b} + \ \frac{A_{1}}{\nu_{1}^{2} - \nu^{2} - i\gamma_{1}\nu} + \frac{A_{2}e^{i\phi}}{\nu_{2}^{2} - \nu^{2} - i\gamma_{2}\nu} \right|
\label{eq:S5}
\end{equation}
Here, \({\widetilde{f}}_{b}\) is a complex-valued non-resonant
background contribution, and each oscillator term describes a cavity
mode with resonance frequency \(\nu_{j}\), linewidth (FWHM)
\(\gamma_{j}\), amplitude \(A_{j}\) and relative phase \(\phi\).

\begin{figure}[htbp]
\centering
\includegraphics[width=0.95\textwidth]{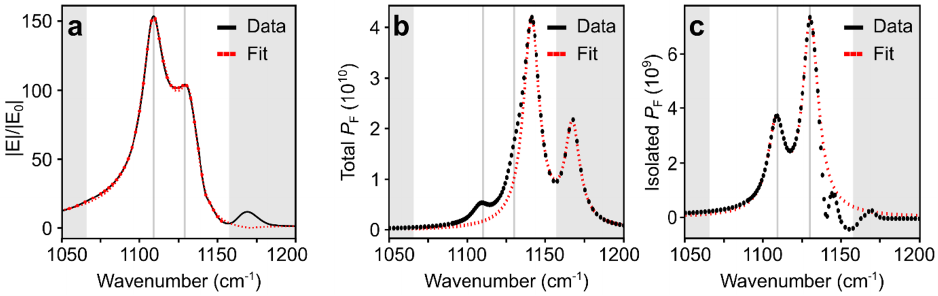}
\caption{\textbf{Extraction of cavity parameters. (a)}
Maximum electric field enhancement \(\max\{|E|(\nu)/|E_{0}|(\nu)\}\) in
the NPoM gap under plane-wave excitation (black) and its corresponding
fit to Eq. S5 (red dashed line), used to extract the resonance
frequencies and linewidths of the NPoM cavity modes. \textbf{(b)} Total
Purcell factor \(P_{F}\) (black), with background contributions fitted
by to Eq. S11 (red dashed line). \textbf{(c)} Isolated Purcell factor
\(P_{F}\) after subtraction of the background (blue line). Intrinsic
NPoM cavity resonances (orange) and their fit to Eq. S11 (orange dashed
line), used to extract linewidths and resonance positions. Grey vertical
lines in all panels mark the resonance frequencies of the \(\mathrm{L}_{01}\) and
\(\mathrm{L}_{02}\) NPoM modes.}
\label{fig:siS5}
\end{figure}

Fitting this model to the electric field in the gap (black curve in \hyperref[fig:main2]{Fig.
2} of the main text) yields the parameters \({\widetilde{f}}_{b} = \)
1.434 + \emph{i}0.01,\(\ \ \phi = \) 3.174 rad, \(\nu_{1} = \) 1108.1
cm$^{-1}$, \(\gamma_{1} = \) 13.7 cm$^{-1}$,
\(\nu_{2} = \) 1132.8 cm$^{-1}$, \(\gamma_{2} = \) 11.8
cm$^{-1}$. In \hyperref[fig:siS5]{Fig.~S5a}, we compare the electric field
enhancement in the NPoM gap (plotted in black) with the corresponding
fit (plotted in red). From the fit values, we obtain \(Q_{1} = \) 81
for the fundamental NPoM cavity mode and \(Q_{2} = \) 96 for the
second-order mode.

To determine the Purcell factor and, consequently, the mode volume, we
perform numerical simulations using a point-dipole source instead of a
plane wave. In particular, the dipolar source must be positioned at the
center of the NPoM gap. This approach provides access to quantities such
as the local density of optical states \(\rho\) (LDOS), the decay rate
of the emitter \(\Gamma\), which characterize light-matter interaction
at the cavity.\cite{snovotny6}

The electromagnetic response of the system to a dipole source is
described through the Green's tensor formalism.\cite{snovotny6} The
electric field generated by a dipole moment \(\overrightarrow{p}\) at
position \({\overrightarrow{r}}_{0}\) \hspace{0pt}is given by
\begin{equation}
\overrightarrow{E}\left( \overrightarrow{r},\omega \right) = \mu_{0}\omega^{2}\overset{\overleftrightarrow{}}{G}\left( \overrightarrow{r},{\overrightarrow{r}}_{0};\omega \right) \cdot \overrightarrow{p}, 
\label{eq:S6}
\end{equation}
where
\(\overset{\overleftrightarrow{}}{G}(\overrightarrow{r},{\overrightarrow{r}}_{0};\omega)\)
is the dyadic Green's tensor of the system, \(\omega = 2\pi\nu\) is the
angular frequency and \(\mu_{0}\) is the vacuum permeability. From this
formalism, the power emitted by the dipole is given by 
\begin{equation}
P(\omega) = \frac{\omega}{2}\, \operatorname{Im}\left\lbrack {\overrightarrow{p}}^{*} \cdot \overrightarrow{E}\left( r_{0},\omega \right) \right\rbrack,
\label{eq:S7}
\end{equation}
which quantifies the coupling between the dipole and the cavity modes
and is directly computed from the local fields obtained in the
simulations. The emitted power is directly related to both the LDOS
\(\rho\) and to the decay rate of the emitter. The LDOS \(\rho\) at the
dipole position is given by
\begin{equation}
\rho\left( \overrightarrow{r},\ \omega \right) = P(\omega)\frac{4\varepsilon_{0}}{\pi\omega^{2}}\frac{1}{\left| \overrightarrow{p} \right|^{2}}\ , 
\label{eq:S8}
\end{equation}
while the emitter decay rate is 
\begin{equation}
\Gamma(\omega) = \frac{P(\omega)}{\hslash\omega}.
\label{eq:S9}
\end{equation}

The Purcell factor is defined as the modification of the emission rate
of an emitter inside of a cavity with respect to its emission in free
space. To compute it, we perform a simulation with the same dipolar
source in vacuum. The Purcell factor is then given by
\begin{equation}
P_{F}(\omega) = \frac{\Gamma(\omega)}{\Gamma_{0}(\omega)} = \frac{P(\omega)}{P_{0}(\omega)}
\label{eq:S10}
\end{equation}

The Purcell factor calculated for the NPoM cavity is shown in black in
\hyperref[fig:siS5]{Fig.~S5b}. In this spectrum, two broad and prominent peaks are observed,
which correspond to the collective contribution of a quasi-continuum of
higher-order modes.\cite{sdelga20147} As a result, the cavity modes of
interest of this study, namely the fundamental \(\mathrm{L}_{01}\) and
second-order \(\mathrm{L}_{02}\) NPoM resonances discussed in the main text,
appear as weaker shoulders superimposed on top of a dominant background
corresponding to contribution of the higher-order
modes.\cite{sesteban20228}

To properly isolate and analyze the cavity modes of interest, the
background must be removed. To this end, the background is fitted using
a model of two Lorentzian functions,
\begin{equation}
f_{2}(\nu) = \ h_{1}\frac{\gamma_{1}}{\left( \nu - \nu_{1} \right)^{2} + \gamma_{1}^{2}} + h_{2}\frac{\gamma_{2}}{\left( \nu - \nu_{2} \right)^{2} + \gamma_{2}^{2}} + b
\label{eq:S11}
\end{equation}

The resulting fit (plotted in dashed-red in \hyperref[fig:siS5]{Fig.~S5b}) is subtracted from
the total Purcell factor (plotted in black in \hyperref[fig:siS5]{Fig.~S5b}). The resulting
background-free Purcell spectrum is shown in blue in \hyperref[fig:siS5]{Fig.~S5c},
exhibiting two clear peaks at the resonance frequencies of the
\(\mathrm{L}_{01}\) and \(\mathrm{L}_{02}\) NPoM modes. To properly extract the linewidths
and the corresponding quality factors of these modes, we fit the
background-free Purcell spectrum using Eq.~\ref{eq:S11}. The fit is shown as an
orange dashed line in \hyperref[fig:siS5]{Fig.~S5c}.

For completeness, we provide in \hyperref[tab:S1]{Table~S1} the resonance frequencies,
spectral linewidths and quality factor of the \(\mathrm{L}_{01}\) and \(\mathrm{L}_{02}\)
NPoM modes. These values are consistently obtained from physical
observables, such as electric field enhancement (\hyperref[fig:siS5]{Fig.~S5a}), the LDOS
(Eq.~\ref{eq:S8}, radiative decay rate (Eq.~\ref{eq:S9} or the isolated Purcell factor
(\hyperref[fig:siS5]{Fig.~S5c}). Overall, we obtain \(Q_{1}\) \(\sim\) 80 for the \(\mathrm{L}_{01}\) mode
and \(Q_{2}\) \(\sim\) 110 for the \(\mathrm{L}_{02}\) mode. This confirms that the
extracted quality factors are independent of the specific observable
used for the fitting, with only slight numerical discrepancies.

\begin{table}[htbp]
\centering
\caption{Resonance frequencies \(\nu_{j}\),
spectral linewidths (FWHM) \(\gamma_{j}\) and quality factors \(Q_{j}\)
for the first (\(j = \) 1) and second (\(j = \) 2) NPoM cavity modes,
obtained from different physical observables such as the electric field
enhancement in the NPoM gap \(|E|/|E_{0}|\), the local density of
optical states (LDOS), the radiative decay rate \(\Gamma\), and the
isolated \(P_{F}\) after background subtraction.}
\label{tab:S1}
\medskip
\begin{tabular}{lcccc}
\hline\hline
 & \(|E|/|E_0|\) & LDOS & \(\Gamma\) & Isolated \(P_F\) \\
\hline
\(\nu_{1}\) (cm$^{-1}$) & 1108 & 1108 & 1108 & 1108 \\
\(\gamma_{1}\) (cm$^{-1}$) & 14 & 12 & 13 & 12 \\
\(Q_{1}\) & 81 & 89 & 86 & 91 \\
\(\nu_{2}\) (cm$^{-1}$) & 1133 & 1129 & 1131 & 1130 \\
\(\gamma_{2}\) (cm$^{-1}$) & 12 & 8 & 10 & 12 \\
\(Q_{2}\) & 96 & 142 & 109 & 97 \\
\hline\hline
\end{tabular}
\end{table}

Once both the quality factor \(Q\) and the Purcell factor \(P_{F}\) are
determined, the mode volume \(V\) can be extracted. The mode volume
quantifies the spatial confinement of the electromagnetic field inside
the cavity. For a resonant cavity supporting well-defined eigenmodes
(\(j = \) 1, 2, \ldots), the Purcell factor of each mode \(j\) is related
to the quality factor \(Q_{j}\ \)and the mode volume \(V_{j}\) as
\begin{equation}
P_{F,j} = \frac{3}{4\pi^{2}}\left( \frac{\lambda_{j}}{n} \right)^{3}\frac{Q_{j}}{V_{j}}, 
\label{eq:S12}
\end{equation}
where \(\lambda_{j}\) is the resonance wavelength of the mode \(j\) and
\(n\) is the refractive index of the reference medium (in our case \(n\)
= 1, as the reference emission is in vacuum). From this relation, the
mode volume of each mode is obtained as
\begin{equation}
V_{j} = \frac{3}{4\pi^{2}}\left( \frac{\lambda_{j}}{n} \right)^{3}\frac{Q_{j}}{P_{F,j}}.
\label{eq:S13}
\end{equation}

Using the extracted values of \(Q_{j}\) (\hyperref[tab:S1]{Supplementary Table~S1}) and the
Purcell factor \(P_{F,j}\) (\hyperref[fig:siS5]{Fig.~S5b}) at each resonance, we obtain mode
volumes of approximately 1500 nm$^{3}$ for the \(\mathrm{L}_{01}\)
mode (\(j = \) 1) and 700 nm$^{3}$ for the \(\mathrm{L}_{02}\) mode
(\(j = \) 2). When normalized to the resonant wavelength of each of the
modes (\(\lambda_{1} = \) 9000 nm,\(\ \lambda_{2} = \) 8850 nm), the mode
volume can be expressed as a dimensionless compression factor, yielding
values of approximately 2 \(\cdot\) 10\textsuperscript{-9} for \(\mathrm{L}_{01}\) and 1
\(\cdot\) 10\textsuperscript{-9} for \(\mathrm{L}_{02}\). The full set of mode volumes
and compression factors for \(\mathrm{L}_{01}\) and \(\mathrm{L}_{02}\) is reported in
\hyperref[tab:S2]{Supplementary Table~S2}.

\begin{table}[htbp]
\centering
\caption{Mode volumes \(V_{j}\) and
wavelength-normalized compression factors \(V_{j}/\lambda_{j}^{3}\) for
the first (\(j = \) 1) and second (\(j = \) 2) NPoM cavity modes, obtained
from the extracted quality factors \(Q_{j}\) and Purcell enhancements
\(P_{F}\) reported in Supplementary Table S1.}
\label{tab:S2}
\medskip
\begin{tabular}{lcccc}
\hline\hline
 & \(|E|/|E_0|\) & LDOS & \(\Gamma\) & Isolated \(P_F\) \\
\hline
\(V_{1}\) (nm$^{3}$) & 1312 & 1458 & 1355 & 1485 \\
\(V_{1}/\lambda_{1}^{3}\) & \(1.8\cdot10^{-9}\) & \(2.0\cdot10^{-9}\) & \(1.9\cdot10^{-9}\) & \(2.0\cdot10^{-9}\) \\
\(V_{2}\) (nm$^{3}$) & 690 & 1040 & 762 & 699 \\
\(V_{2}/\lambda_{2}^{3}\) & \(1.0\cdot10^{-9}\) & \(1.5\cdot10^{-9}\) & \(1.1\cdot10^{-9}\) & \(1.0\cdot10^{-9}\) \\
\hline\hline
\end{tabular}
\end{table}

\newpage

\subsection{Modification of the modes upon tip approach}
\label{si:section-6-modification-of-the-modes-upon-tip-approach}

In the main text, we show that the presence of the tip leads to a
stronger enhancement of the local electric field in the NPoM gap, while
the spectral position and linewidth of the cavity resonances remain
essentially unchanged. However, given the strong near-field interaction
between the tip and the NPoM cavity, it is not obvious a priori that the
tip does not perturb the intrinsic NPoM cavity properties. In this
section, we verify this by quantifying the effect of the tip on the
quality factor and on the lateral confinement of the NPoM cavity modes.

\begin{figure}[htbp]
\centering
\includegraphics[width=0.75\textwidth]{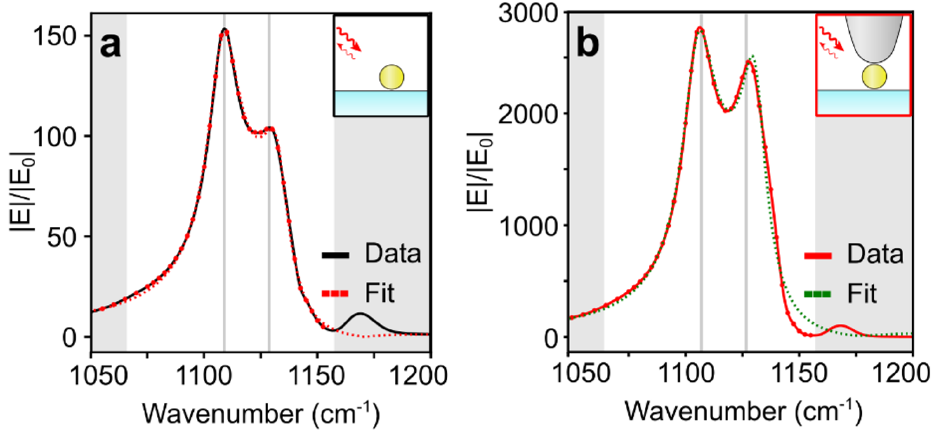}
\caption{\textbf{Variation of \(\mathbf{Q}\)
upon tip approach.} Maximum field enhancement
\(\max\{|E|(\nu)/|E_{0}|(\nu)\}\) in the NPoM gap under plane-wave
excitation (solid line) and its corresponding fit to a double Lorentz
harmonic oscillator model (dashed line), used to extract the resonance
frequencies and linewidths of the NPoM cavity modes. \textbf{(a)} NPoM
cavity in absence of the tip \textbf{(b)} NPoM cavity in presence of the
tip.}
\label{fig:siS6}
\end{figure}

To evaluate the quality factor \(Q\) we follow the same procedure
described in the previous section, where \(Q\) is extracted from the
electric field spectra in the NPoM gap. In \hyperref[fig:siS6]{Fig.~S6} we compare the
spectral response of the system without and with the tip, corresponding
to the data shown in \hyperref[fig:main2]{Fig.~2b} of the main text. By fitting the electric
field in the presence of the tip (\hyperref[fig:siS6]{Fig.~S6b}) with a double harmonic
oscillator model (Eq.~\ref{eq:S4}, we obtain for the first NPoM cavity mode a
resonance frequency \(\nu_{1} = \) 1105 cm$^{-1}$ and a
linewidth (FWHM) of \(\gamma_{1} = \) 14 cm$^{-1}$, while
for the second-order mode we obtain \(\nu_{2} = \) 1131
cm$^{-1}$, \(\gamma_{2} = \) 11 cm$^{-1}$.
These values correspond to quality factors \(Q_{1} = \) 80 and
\(Q_{2} = \) 108, respectively. When comparing these values with those
obtained for the NPoM cavity in absence of the tip 
(\hyperref[tab:S1]{Supplementary Table~S1}), 
we find a good quantitative agreement. This indicates that the tip
does not introduce additional losses comparable to the non-radiative
losses of the cavity modes nor significantly modify their resonance
frequencies.

\begin{figure}[htbp]
\centering
\includegraphics[width=0.95\textwidth]{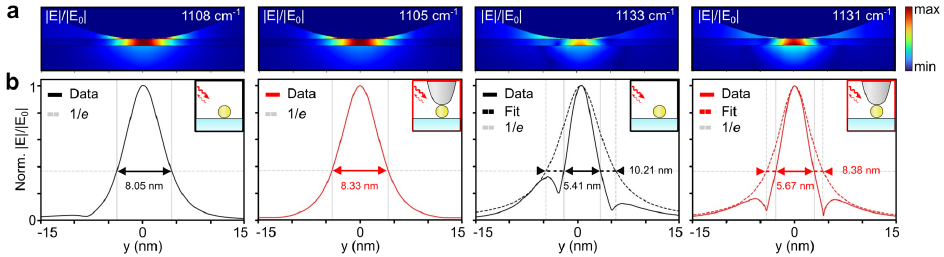}
\caption{\textbf{Lateral confinement of NPoM cavity
modes. (a)} Spatial distribution of the total electric field enhancement
\(|E|(\nu)/|E_{0}|(\nu)\) in the NPoM gap at the resonance frequencies
in presence and absence of the tip. \textbf{(b)} Line profiles of the
normalized electric field amplitude in the gap (solid lines) and the
corresponding Gaussian fits for the second-order mode. Horizontal dashed
lines mark the 1/\(e\) value, from which the spatial extent is defined,
as indicated by the horizontal arrows.}
\label{fig:siS7}
\end{figure}

To quantify the effect of the tip on the lateral confinement of the NPoM
cavity modes, we analyze the spatial distribution of the electric field
in the gap at the resonance frequencies. To this end, we consider the
total field amplitude in the gap region (\hyperref[fig:siS7]{Fig.~S7a}) and extract a line
profile along the y-axis (\(x = \) 0) through the gap horizontal midplane
(\hyperref[fig:siS7]{Fig.~S7b}). The field is normalized to its maximum value, and the
spatial extent \(w\) is defined as the distance between the spatial
positions at which the field amplitude decreases to 1/\(e\) of its
maximum.

We plot the spatial profiles and the corresponding spatial extents \(w\)
in \hyperref[fig:siS7]{Fig.~S7b}. The fundamental \(\mathrm{L}_{01}\) mode exhibits a well-defined
Gaussian shape with a spatial extent \(w = \) 8 nm and a variation of
approximately 3 \% upon tip approach. For the second-order \(\mathrm{L}_{02}\)
mode, the presence of side lobes complicates the analysis. To include
their contribution, we estimate \(w\) using a Gaussian fit (dashed line
in \hyperref[fig:siS7]{Fig.~S7b}) of the full field profile. The extracted spatial extent is
\(w = \) 10.2 nm in absence of the tip and \(w = \) 8.4 nm in presence of
the tip, corresponding to a 18 \% variation. When considering only the
central lobe, the variation upon tip approach is of about 5 \%. In
absolute numbers, the spatial extent of the NPoM modes always remains
\(\sim\) 10 nm.

Overall, these results demonstrate that the presence of the tip has a
negligible impact on both the quality factor and the spatial confinement
of the NPoM modes. They validate the conclusion presented in the main
text: the tip strongly enhances the local field in the NPoM cavity
without significantly altering the intrinsic spectral properties or
spatial extent of the cavity modes, thereby acting primarily as a
near-field probe.

\newpage

\subsection{Calculated demodulated near-field spectra on quartz}
\label{si:section-7-calculated-demodulated-near-field-spectra-on-quartz}

As illustrated in \hyperref[fig:main2]{Fig.~2d} of the main text and 
\hyperref[si:section-2-spectra-as-a-function-of-tipquartz-separation]{Supplementary
Section 2}, 
the tip--quartz distance strongly influences the spectral
position and intensity of the surface phonon polariton. Increasing the
tip--quartz separation by only 10 nm results in a substantial blueshift
and attenuation of the fundamental mode, while the higher-order modes
become strongly suppressed. Since nano-FTIR measurements rely on
modulation of the near-field signal through tip oscillation,
understanding how these resonances evolve as a function of tip--quartz
distance is essential for interpreting the measured spectra.

\begin{figure}[htbp]
\centering
\includegraphics[width=0.9\textwidth]{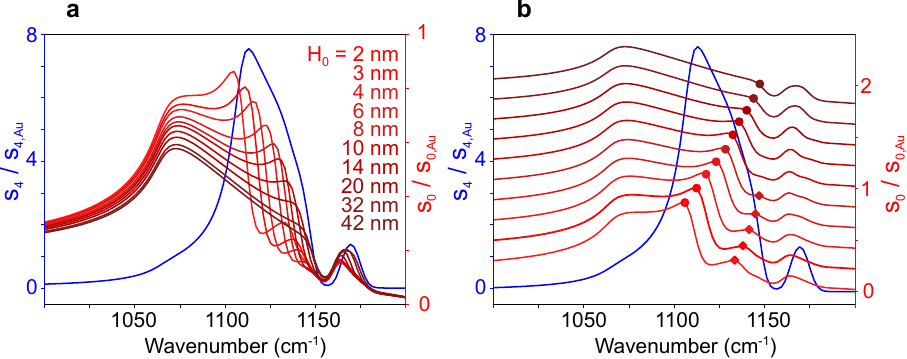}
\caption{\textbf{(a)} Numerical calculation of
demodulated signal (blue) for a tip oscillating with 40 nm tapping
amplitude on top of quartz, compared to non-demodulated signals with a
static tip at different distances \(H_{0}\) from quartz (red). \textbf{(b)} Same
data as in \textbf{(a)}, a vertical offset is applied for better visualization.
Red dots and diamonds mark the spectral position attributed to the
\(\mathrm{L}_{01}\) mode and the \(\mathrm{L}_{02}\) mode respectively.}
\label{fig:siS8}
\end{figure}

To investigate the effect of the dynamically varying tip--quartz
separation during a tip tapping cycle, we employ the spheroid tip
model.\cite{svoronin20259} We first calculate non-demodulated spectra
corresponding to a static tip positioned at fixed distances above the
quartz surface (\hyperref[fig:siS8]{Fig.~S8}, red curves). These calculations are performed
for tip--quartz separations \(H_{0}\) ranging from 2 to 42 nm and are
referenced to a gold substrate. For the smallest separation (\(H_{0} = \)
2 nm), the fundamental \(\mathrm{L}_{01}\) mode appears at 1105
cm$^{-1}$. As the tip is retracted, this resonance exhibits
a pronounced blueshift, reaching 1146 cm$^{-1}$ at
\(H_{0} = \) 42 nm, while its amplitude decreases due to the reduced
electromagnetic interaction between the tip and the quartz substrate. A
similar trend is observed for the \(\mathrm{L}_{02}\) mode, which is located at
1134 cm$^{-1}$ for \(H_{0} = \) 2 nm. This mode also
blueshifts with increasing distance but decays much faster in amplitude,
becoming negligible for \(H_{0}\) \textgreater{} 10 nm.

We next calculate the demodulated near-field response using the same
model. Specifically, we evaluate the fourth harmonic of the scattered
field for a tip oscillating with a tapping amplitude of 40 nm and a
minimum tip--quartz distance of 2 nm. The resulting spectrum (\hyperref[fig:siS8]{Fig.~S8},
blue curve), also referenced to gold, exhibits a substantially broader
resonance than the corresponding static spectra. This broadening arises
from the demodulation process. During tapping-mode operation, the tip
samples a continuous range of tip--quartz distances within each
oscillation cycle, and the detected higher-harmonic signal represents a
weighted average of the near-field response over this trajectory. Since
the resonance frequency of the L\textsubscript{01} mode depends strongly
on the tip--quartz distance 
(see \hyperref[si:section-2-spectra-as-a-function-of-tipquartz-separation]{Supplementary
Section 2}), 
averaging over the oscillation cycle results in a broadening of the main spectral
peak, while other peaks remain indistinguishable. This behavior is
consistent with the experimental results (see inset of \hyperref[fig:main1]{Fig.~1d} in the
main text).

In summary, direct nano-FTIR measurements on quartz are not suitable for
accessing the phononic cavity modes that emerge at nanometer-scale
tip--quartz separations. In contrast, in the NPoM geometry the
nanoparticles remain at a fixed distance from the quartz substrate,
enabling the cavity modes to persist and be directly observed in the
measured spectra. 

\newpage

\subsection{Experimental Methods}
\label{si:section-8-experimental-methods}

\textbf{Sample preparation:} We use alpha-quartz substrates with a
{[}0001{]} orientation (CrysTek, c-cut, citrate capping agent). We
prepare the samples via dropcasting, using commercially available Au
nanoparticles (40 nm, 60 nm and 80 nm gold colloids from BBI Solutions).
After increasing the nanoparticle solution concentration by a factor of
four via centrifugation (5 minutes with 10000 rpm for 40 nm, 6000 rpm
for 60 nm and 5000 rpm for 80 nm). We deposit a drop of the solution on
the quartz substrates. The samples are then left in a fume hood to allow
most of the water to evaporate. This process takes 30 -- 60 minutes,
depending on the drop size and conditions. The sample is then gently
rinsed with distilled water to remove residual impurities, followed by
5-minute of plasma cleaning.

\textbf{FTIR measurements:} FTIR spectra are recorded with a Bruker
Hyperion 2000 FTIR spectrometer using a 15x IR Cassegrain objective with
numerical aperture \(\mathrm{NA} = \) 0.4. The measurements are performed in
reflection mode with unpolarized light. We use a separate gold substrate
as a reference. The spectrum shown in \hyperref[fig:siS1]{Fig.~S1} is an average of 200
spectra recorded over the spectral range 600 -- 4000
cm$^{-1}$, with a resolution of 4 cm$^{-1}$.

\textbf{Nano-FTIR measurements:} We use a commercial nano-FTIR setup
(Attocube Systems AG), in which an oscillating Pt/Ir-coated AFM tip
(nano-IR, NanoWorld AG), operated at a frequency at a frequency
\(\Omega\) \(\approx\) 250 kHz with a tapping amplitude of 70 nm, is illuminated
by p-polarized mid-IR broadband radiation generated by a supercontinuum
laser (Femtofiber pro IR and SCIR; Toptica; average power of 0.5 mW;
wavenumber range 800 -- 1400 cm$^{-1}$). The detector signal
is demodulated at a frequency \(n\Omega\) (with \(n = \) 3, 4) for
effective background suppression. Interferograms are measured by
recording the demodulated detector signal as a function of the position
of the reference mirror and averaged over 20 runs. For apodization of
the interferograms, a Tukey window function with \(\alpha = \) 0.2 is
applied. After zero-filling (3\(\times\) padding), we Fourier transform the
interferograms to obtain complex-valued near-field point spectra,
\(E_{s}(\omega)\). We normalize the resulting spectra to a reference
spectrum measured on a separate gold sample, \(E_{s,\mathrm{Au}}(\omega)\),
acquired under the same conditions and parameters. Each interferogram
consists of 1024 pixels, with an integration time of 15 ms per pixel, so
that each measurement lasted 5 minutes. The spectral resolution is 6.25
cm$^{-1}$, which is the limit of the setup.

\textbf{Retraction curve measurements:} We used the same setup and
settings as described in the section above. The tip is illuminated with
radiation from a quantum cascade laser (QCL; MIRcat; Daylight Photonics)
at wavenumbers between 1050 and 1150 cm$^{-1}$. The
detector signal is demodulated at a frequency 4\(\Omega\) for effective
background suppression. Each measurement consists of a retraction scan
in which the sample is retracted in 1000 steps over a total distance of
150 nm, with an integration time of 50 ms per step.

\subsection{Numerical Methods}
\label{si:section-9-numerical-methods}

\textbf{IR materials response:} We model the IR response of the
materials considered in the numerical calculations as follows. The
dielectric function for the platinum (Pt) tip, corresponds to a Brendel
and Bormann model.\cite{selazar199810} The dielectric function of gold
(Au) is obtained by linearly interpolating data from
literature.\cite{sbabar201511} For the quartz substrate, we obtain the
in-plane dielectric function \(\varepsilon_{\bot}(\omega)\) by fitting
FTIR reflectivity measurements described in \hyperref[si:section-1-far-field-reflectivity-measurements-on-quartz]{Supplementary Section 1}. For
simplicity, we neglect the intrinsic optical anisotropy of quartz and
treat its optical response as isotropic with dielectric tensor
\(\ \overset{\overleftrightarrow{}}{\varepsilon} = \mathrm{Diag}(\varepsilon_{\bot}\ ,\varepsilon_{\bot}\ ,\varepsilon_{\bot}).\)
The impact of this approximation is discussed in the 
\hyperref[si:section-3-effect-of-quartz-anisotropy-on-the-local-field-enhancement-in-the-npom-cavity]{Supplementary Section 3}.

\textbf{Numerical simulations:} The electromagnetic simulations are
performed with a commercial FEM package (COMSOL). The simulation
universe contains a Pt tip, modeled as a 1 micron height half ellipsoid
with either 25 nm or 50 nm apex diameter, a spherical gold nanoparticle
with no facet of diameter \(d_{\mathrm{NP}} = \) 40, 50, 60, 80 nm, which is
separated from a 1 um thick quartz bulk by a 1 nm thin dielectric layer
with refractive index \(n = \) 1.4. The simulation universe consists of a
box of 6.5 $\mu$m width and 4.5 $\mu$m height, surrounded by a 3 $\mu$m thick PML
layer. Special care is taken to mesh the near-contact regions of the
tip-enhanced phononic NPoM, that is the tip-NP and NP-quartz gaps. From
the numerical calculations, we obtain the maximum of the scattered field
in the horizontal midgap plane between the NP and substrate. We
illuminate the system with a p-polarized plane-wave excitation incident
at an angle of $35^\circ$ from the quartz surface (\(x - y\) plane). 

\newpage


\end{document}